\documentclass[12pt, oneside, reqno]{amsart}
 \usepackage{graphicx}
\usepackage{amsmath,amsthm,amsfonts,amscd,amssymb,mathrsfs,hyperref}
\usepackage[all]{xypic}
\usepackage{amscd}   
\usepackage[all]{xy} 
\usepackage{booktabs} 
\usepackage{physics}
\usepackage{hyperref}
\hypersetup{colorlinks=true,citecolor=blue, urlcolor= cyan}
\usepackage{qcircuit}

\usepackage{multicol}
\usepackage{multirow}
\usepackage{bm}

\usepackage{ytableau}
\usepackage[enableskew]{youngtab} 

\usepackage[backend=biber,
style=numeric,
sorting=none
]{biblatex}

\usepackage{nicefrac}
\usepackage{xfrac}

\usepackage{adjustbox}
\usepackage{booktabs, makecell, multirow}
\usepackage{geometry}
\usepackage{array}
\usepackage{tabularx}

\usepackage{color}

\usepackage{tikz}

\makeatletter
\newtheorem{rep@theorem}{\rep@title}
\newcommand{\newreptheorem}[2]{%
\newenvironment{rep#1}[1]{%
 \def\rep@title{#2 \ref{##1}}%
 \begin{rep@theorem}}%
 {\end{rep@theorem}}}
\makeatother

\newtheorem{theorem}{Theorem}[section]
\newreptheorem{theorem}{Theorem}
\newreptheorem{lemma}{Lemma}

\def\phi{ \varphi }

\theoremstyle{definition}

\theoremstyle{remark}
\newtheorem{remark}[theorem]{Remark}

\newcommand{\ccirc}[1]{\xymatrix@1{+<1ex>[o][F-]{#1}}}

\title[Two Qubit Games on NISQ Computers]{Implementing 2-qubit pseudo-telepathy games on noisy intermediate scale quantum computers}
\author{Colm Kelleher, Mohammad Roomy, Fr\'ed\'eric Holweck}
\address{Université Bourgogne Franche-Comté, Laboratoire interdisciplinary Carnot de Bourgogne UMR6303, 25000 Dijon, FRANCE}
\email{colm.kelleher@utbm.fr}
\address{Université Bourgogne Franche-Comté, Laboratoire interdisciplinary Carnot de Bourgogne UMR6303, 25000 Dijon, FRANCE}
\email{mohammad\_roomy@etu.u-bourgogne.fr}
\address{University of Technology of Belfort-Montbéliard, Laboratoire interdisciplinary Carnot de Bourgogne UMR6303, ICB-UTBM, 90000 Belfort, FRANCE}
\address{Mathematics and Statistics Dept, Aubrun University, Auburn, AL, USA}
\email{frederic.holweck@utbm.fr}

\begin{document}

\maketitle

\begin{abstract}
It is known that Mermin-Peres like proofs of quantum contextuality can furnish non-local games with a guaranteed quantum strategy, when classically no such guarantee can exist. This phenomenon, also called quantum pseudo-telepathy, has been studied  in the case of the so-called Mermin Magic square game. In this paper we review in detail two different ways of implementing on a quantum computer such a game and propose a new Doily game based on the geometry of 2-qubit Pauli group. We show that the quantumness of these games are almost revealed when we play them on the IBM Quantum Experience, however the inherent noise in the available quantum machines prevents a full demonstration of the non-classical aspects.   
\end{abstract}

\section{Introduction}\label{sec:intro}
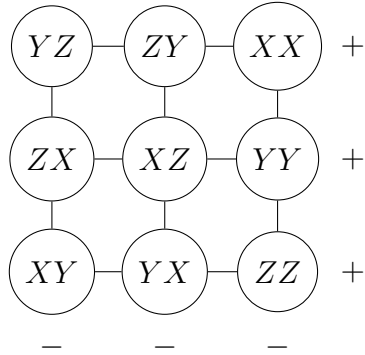
\begin{figure}[h!]
\centering
\begin{tikzpicture}[
operator/.style={circle, draw, minimum size=1.0cm},
]
\foreach \x/\y/\num in {1/3/$YZ$, 2/3/$ZY$, 3/3/$XX$, 1/2/$ZX$, 2/2/$XZ$, 3/2/$YY$, 1/1/$XY$, 2/1/$YX$, 3/1/$ZZ$}
    \node[operator] (\x-\y) at (1.5*\x, 1.5*\y) {\num};
    
\draw[] (1-1.east) -- (2-1.west);
\draw[] (2-1.east) -- (3-1.west);
\draw[] (1-2.east) -- (2-2.west);
\draw[] (2-2.east) -- (3-2.west);
\draw[] (1-3.east) -- (2-3.west);
\draw[] (2-3.east) -- (3-3.west);

\draw[] (1-1.north) -- (1-2.south);
\draw[] (1-2.north) -- (1-3.south);
\draw[] (2-1.north) -- (2-2.south);
\draw[] (2-2.north) -- (2-3.south);
\draw[] (3-1.north) -- (3-2.south);
\draw[] (3-2.north) -- (3-3.south);

  \foreach \x in {1,2,3}
    \node at (1.5*\x, 0.5) {$-$};

  \foreach \y in {1,2,3}
    \node at (5.5, 1.5*\y) {$+$};

\end{tikzpicture}
\caption{An example of a Mermin-Peres Magic square: The nodes represent two-qubit Pauli observables with the short handed notation $AB=A\otimes B$. Observables on the same lines commute, non-colinear observables anti-commute.}
\label{fig:merminsquare}
\end{figure}

Quantum contextuality is a counter-intuitive feature of quantum physics that is now considered as a resource for quantum computation. See \cite{Budroni21} for a recent and comprehensive survey on the subject.  Among the various presentations of  the concept, state independent formulations can be established as configurations of observables. The most famous one being the Mermin-Peres square \cite{Mermin93,Peres90}. One instance of this square of observables is given in Figure \ref{fig:merminsquare}.

The nodes correspond to two-qubit Pauli observables, the tensor product is omitted, and the edges of the grid correspond to set of mutually commuting observables whose product is $\pm I_d$. The sign is indicated on Figure \ref{fig:merminsquare} for each row and column. Such a set of mutually commuting observable whose product is $\pm I_d$ will be called a context. The possible outcomes of the measurements for each node of the square is $\pm 1$ as they are the eigenvalues of two-qubit Pauli matrices. Moreover according to the rules of Quantum Mechanics (QM), the product of the eigenvalues for a set of mutually commuting observables has to be equal to an eigenvalue of the product of observables, i.e. the contexts impose global constraints on the outcomes of the measurements. For instance the outcomes of the measurements on the first row will be randomly $\pm 1$ but their product should be $+1$. Now a simple checking argument shows that there is no classical deterministic function that can assign $\pm 1$ to each node and at the same time satisfy all the constraints imposed by the rows and columns of the square. The only possibility would be to have a classical model whose measurement outcomes change depending on the context (set of mutually commuting observables) that is considered. One therefore says that there is no Non Contextual Hidden Variables model that can reproduce the outcomes of QM. This statement is equivalent to the famous Kochen-Specker (KS) Theorem \cite{Kochen_specker,Bell_HV}.

In \cite{Aravind02,Aravind04}, Aravind proposed a non-local quantum game based on the Mermin's magic square, promoting by this the Mermin-like proof of KS into a proof of Bell's Theorem. A non-local quantum game is a game where players have an advantage to play with quantum resources against classical players. The game is, among other pseudo-telepathy games, presented in \cite{Brassard05} where a circuit version of game is provided. In \cite{Arkhipov12} a geometrical criteria is given to characterize and extend the Magic square game to other scenario and in \cite{Fialik12} the behaviour of the game under noise is investigated. Xu et. al \cite{xu_experimental_2022} implemented the game in a photonic quantum circuit, demonstrating the quantum phenomena under investigation. \\

In this paper we implement different Mermin-like games on a Noisy Intermediate Scale Quantum (NISQ) computer available online \cite{ibmq}. These games are implemented using multiple methods, and compared against idealised and noisy simulators. Whilst the games demonstrate the quantum advantage in the simulators - winning the game at a higher rate than any classical strategy allows - the results on the actual NISQ computers fall short, due to circuit topology-induced noise.\\
The paper is organized as follows. In Section \ref{sec:Quantum_games} we introduce the notion of quantum games, recalling first the principle of the Mermin Magic square game before introducing the Doily game, which is based on a larger configuration that encodes the commutation relations of all two-qubit Pauli observables. In Section \ref{sec:ibmtest} we detail how these games are implemented in a quantum circuit, including various scenarios and methodologies. Finally we report our experimental tests on the IBM Quantum Experience in Section \ref{sec:results}, comparing them to classical bounds and simulations. Section \ref{sec:conclusion} ends the paper with concluding remarks.

\section{Quantum Games}\label{sec:Quantum_games}
\subsection{The Mermin-Peres Magic Square game}\label{sec:Mermin}
Let us start by introducing the scenario of the Mermin Magic square game. We will in fact work with two different scenarios. Note that the Mermin's square has the incidence structure of a grid made of $9$ vertices and $6$ edges. This geometric configuration is called a generalized quadrangle because the configuration is triangle free (i.e. it is not possible to connect three non-colinear vertices by less than four edges). A general quadrangle with $s+1$ points/vertices per line/edge and $t+1$ lines/edges through a point is denoted by $GQ(s,t)$. Thus we will denote the incidence structure of the Mermin's square by $GQ(2,1)$. 
\subsubsection{Scenario 1 (The line-line scenario)}
Suppose Alice and Bob are playing against a referee Charlie. Alice and Bob can establish a strategy before the game starts but can not communicate during the game. The game proceed in the following steps:
\begin{itemize}
 \item Charlie sends a question $x\in \{1,2,3\}$ to Alice and a question $y\in \{1,2,3\}$ to Bob.
 \item Alice and Bob send back triplets of numbers $(a_1,a_2,a_3)$ for Alice and $(b_1,b_2,b_3)$ for Bob such that $a_i,b_j\in \{-1,1\}$ with $a_1a_2a_3=1$ and $b_1b_2b_3=-1$. These constraints are called the parity conditions.
 \item Alice and Bob win if and only if $a_y=b_x$.
\end{itemize}
Table \ref{tab:classical} shows a classical strategy that can be used by Alice and Bob to win the game in $8$ out of the $9$ different questions and there is no better classical strategy \cite{Aravind04}. We will denote by $\omega_i(G)$ the classical probability of winning the game $G$ where $G$ is the geometry corresponding to the game. The subscript allows to distinguish the different scenarios. Thus for this first version ($LL$) of the Mermin game we have:
\begin{equation}
 \omega_{LL}(GQ(2,1))=0.\overline{8}.
\end{equation}

\begin{table}[!h]
 \begin{tabular}{|c|r|r|r|c|}
  \hline
  $x\setminus y$ & $1$ & $2$ & $3$ &  Parity\\
  \hline 
  $1$ & $1$ & $1$ &$1$ & $\bf{+}$ \\
  \hline
  $2$ &$-1$ &$1$ &$-1$ & $\bf{+}$\\
  \hline
  $3$ & $1$ &$-1$ & ?? & $\bf{+}$\\
  \hline
   Parity & $\bf{-}$ & $\bf{-}$  &$\bf{-}$& \\
   \hline
 \end{tabular}
 \caption{An example of classical strategy: Alice and Bob share this table before the game begins. The triplet sent back by Alice corresponds to the row for a given $x\in\{1,2,3\}$ while the triplet sent back by Bob corresponds to the column $y\in\{1,2,3\}$. They win when the value given by each player at the intersection point agrees. With this strategy they win the game in $8$ cases out of $9$.}\label{tab:classical}
\end{table}

This game is often described as \textit{pseudo-telepathic} as if Alice and Bob could communicate in a non-classical manner once the game is afoot, they can organise their responses to win with probability $1$. However there is a quantum strategy with no communication that allows Alice and Bob to win the game with such a probability. Suppose Alice and Bob decide before the game starts to share the following four-qubit entangled state:
\begin{equation}\label{eq:state}
 \ket{\psi_{AB}}=\ket{EPR}_{13}\otimes \ket{EPR}_{24}=\dfrac{1}{2}(\ket{0000}+\ket{0101}+\ket{1010}+\ket{1111})
\end{equation}
Alice gets the first two qubits while Bob takes the last two ones.
 To win the game Alice will measure her two qubit state according to the context of observables given by the row $x$ of the Mermin magic square (effectively replacing Table \ref{tab:classical} with Fig. \ref{fig:merminsquare}). Bob will measure his two qubit pair by following the $y$ column context of the square. By performing these sequential measurements they both get a triplet of values $(a_1,a_2,a_3)$ and $(b_1,b_2,b_3)$. The parity conditions will be automatically satisfied because of the constraints imposed by the rows and columns of the magic square. \\
 \\
 One must check to see if the resultant measurements will always agree on the state $\ket{\psi_{AB}}$ when both Alice and Bob consider the same observable $\mathcal{O}_i$, i.e. when the row and column intersect. For any two-qubit operator $\mathcal{O}_{i}$ consisting of an even number of $Y$s (hereby defined as \textit{symmetric}), one can see\cite{Aravind02} that
 \begin{equation}\label{eq:same_eigvalue_sym}
     \mathcal{O}^A_i\otimes \mathcal{O}^B_i \ket{\psi_{AB}}=\ket{\psi_{AB}}
 \end{equation}
 
 Proving that $\ket{\psi_{AB}}$ is an eigenstate of eigenvalue $+1$. Thus, the measurement outcomes for $\mathcal{O}^A_i$ and $\mathcal{O}^B_i$ must agree.\\
 For any operators consisting of an odd number of $Y$s (\textit{skew}), one has that $\ket{\psi_{AB}}$ is an eigenstate of eigenvalue $-1$, and the measurements will always disagree. In this scenario, whenever there is a skew operator, one player (say Alice) always flips her measurement result in order to get agreement with Bob's. One can check from the magic square that this does not affect the overall parity constraints. Under this strategy, Alice and Bob will always return an agreeing result for a common measurement $\mathcal{O}_{i}$, guaranteeing victory for any questions $x,y$ asked by Charlie.

 \subsubsection{ Scenario 2 (The point-line scenario)} There is an alternative version of the game described in \cite{Arkhipov12} and which goes as follows. Like in the first scenario Alice and Bob play against Charlie.
 \begin{itemize}
  \item Charlie sends a question $c\in \{1,2,3,4,5,6\}$ to Bob corresponding to a line of the grid $GQ(2,1)$ and a question $d\in \{1,2,3,4,5,6,7,8,9\}$ to Alice corresponding to a vertex. This vertex should belong to the context $c$ sent to Bob. 
  \item Bob send back a triplet $(b_1,b_2,b_3)$ such that $b_i\in \{-1,1\}$ with the constraints $b_1b_2b_3=1$ if $c\in\{1, 2, 3\}$ and $b_1b_2b_3=-1$ if $c\in\{4,5,6\}$. Alice sends back $a\in\{1,-1\}$.
  \item Alice an Bob win the game if and only if $a=b_d$ where $b_d$ is Bob's element of the triplet $(b_1,b_2,b_3)$ which occupies the vertex $d$ in the grid $GQ(2,1)$.
 \end{itemize}
 
 In this scenario the classical probability of winning the game is higher than in the classical Mermin square game. Indeed there are now $18$ different questions that Charlie can ask and one sees from Table \ref{tab:classical} that Alice and Bob can agree on a classical strategy that works for all questions but one, i.e.
 \begin{equation}
  \omega_{PL}(GQ(2,1))=\dfrac{17}{18}= 0.9\overline{4}
 \end{equation}

The quantum strategy, which is the same as in Scenario $1$ allows Alice and Bob to win the game with probability $1$.

\begin{remark}
 Despite the fact that $\omega_{PL}(GQ(2,1))>\omega_{LL}(GQ(2,1))$ there are two motivations to look at this alternative scenario. First as pointed out in \cite{Arkhipov12} it applies for all point-line geometries (called arrangements in \cite{Arkhipov12}) even if the arrangement does not have a canonical partition of lines that can be divided in two sets of lines  which do not intersect within each set (like the rows and columns of $GQ(2,1)$). The second motivation is in terms of implementation. As we will see in Section \ref{sec:ibmtest} this second scenario leads us an alternative implementation of the game that allows us to optimize the resources provided by the online quantum computers.
\end{remark}

\subsection{The Doily Game}\label{sec:doily}

\begin{figure}[!h]
 \begin{center}
  \includegraphics[width=7cm]{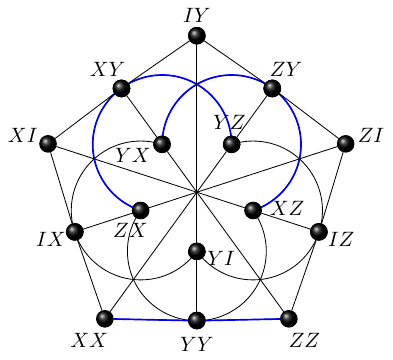}
  \caption{The Doily: A $15_3$ point-line configuration that encodes the commutation relations of the two-qubit Pauli group. $I$ denotes the 1-qubit identity operator, and as before tensor products are omitted. The blue lines are such that the product of the observable is $-I_{4}$ while it is $+I_{4}$ for the other lines.}\label{fig:doily}
 \end{center}
\end{figure}

The doily is another point-line configuration, like the grid, that can be labelled by two qubit observables such that observables on the same line are mutually commuting and form a context. In fact this $15_3$ point-line configuration, $15$ points, $15$ lines, $3$ points per line, $3$ lines per point, encapsulates the commutation relations of the two qubit Pauli group. The Figure \ref{fig:doily} provides a labelling of the Doily by two qubit Pauli observables such that the observables commute if they are on the same line and do not commute otherwise.

The Doily is a generalisation of the magic Mermin square, in that it encodes the commutation relations of \textit{all} non trivial two-qubit Pauli observables, and thus contains the magic square as a subgeometry within it.\\
It is known that the Doily is a contextual configuration of observables \cite{cabello_proposed_2010} and its contextuality has been tested experimentally on an online quantum computer in \cite{holweck_testing_2021}. Like for the Mermin magic square, some lines are negative, more precisely $3$ of them. As a point line geometry one can check that the doily is the generalized quadrangle $GQ(2,2)$.

Let us consider the following two scenarios based on the incidence structure of $GQ(2,2)$.

\subsubsection{Scenario 1 (the line-line scenario)}
To built a scenario where the questions sent by the Referee correspond to lines of $GQ(2,2)$ Charlie has to select couples of lines of $GQ(2,2)$ that intersect. The geometry being made of $15$ lines there are $15\times 6=90$ different questions that Charlie can send to Alice and Bob. We also suppose that the vertices are ordered on each lines of the geometry so it makes sense to talk about the position of a vertex on a given line. The game can be described as follows:
\begin{itemize}
 \item Charlie sends a question $l_1$ to Alice and $l_2$ to Bob where $l_1$ and $l_2$ correspond to two (ordered) intersecting lines of $GQ(2,2)$.
 \item Alice and Bob send back triplets $(a_1,a_2,a_2)$ and $(b_1,b_2,b_3)$ to Charlie with $a_i, b_j \in \{-1,1\}$ and $a_1a_2a_3=1/-1$ if $l_1$ is a positive/negative line, $b_1b_2b_3=1/-1$ if $l_2$ is a positive negative line.
 \item Alice and Bob win the game if $a_x=b_y$ where $x$ and $y$ are the positions in $l_1$ and $l_2$ corresponding to the intersection of the two lines.
\end{itemize}

It is known that there is no classical function that label the vertices of the doily by $\pm 1$ and satisfy the signs conditions imposed by the Doily. In fact out of the $15$ constraints, only $12$ can be satisfied classically for instance by labeling all vertices by $+1$. This is what proves that the doily is a contextual configuration \cite{cabello_proposed_2010,holweck_testing_2021}. Suppose that for all positive lines Alice and Bob send back a triplet of $+1$ and for a given negative line they randomly choose in which position they assign $-1$. Then they will classically win with probability $1$ if they both got two positive lines and they win with probability $2/3$ if one of them has a positive line and the other a negative line. As two negative lines do not intersect, the case where they both get a negative line does not occur.

\begin{equation}
    \omega_{LL}(GQ(2,2)) = 1 - \frac{9}{15}\times\frac{2}{3}\times\frac{1}{3}=\frac{13}{15}= 0.8\overline{6}
\end{equation}
The quantum strategy is analogous to the one of the Mermin game, using the same shared state $\ket{\psi_{AB}}$ from \eqref{eq:state} and the same skew-flipping protocol. As in the Mermin game, the quantum strategy for the Doily game ensures Alice and Bob win with probability $1$.

\subsubsection{Scenario 2 (the point-line scenario)}
Let us now consider the point-line scenario associated to the incidence structure of the doily.
\begin{itemize}
 \item Charlie sends a question corresponding to a point $p$ of the Doily to Alice and a question $l$ corresponding to a vertex of the Doily to Bob such that $p\in l$ and the point of the line are ordered.
 \item Alice sends back $a\in \{1,-1\}$ and Bob sends back a triplet $(b_1,b_2,b_3)$ such that $b_i \in \{1,-1\}$ and $b_1b_2b_2=\pm 1$ according to the sign constrain given by the line $l$ of the doily.
 \item Alice and Bob win the game if $a=b_x$ where $x$ is the position of $p$ in $l$.
\end{itemize}

Again as the contextual nature of the doily insures us that one can not satisfy more than $12$ constraints with a classical function, a classical strategy consists for example for Alice to always return $+1$ and for Bob to return triplets of $+1$ when he got a positive line and then to randomly choose one element of his triplet to be negative if he got a negative line. Thus one obtains in this case,
\begin{equation}
    \omega_{PL}(GQ(2,2))=1-\frac{3}{15}\times\frac{1}{3}=\frac{14}{15}=0.9\overline{3}
\end{equation}

Like in Scenario $1$ a quantum strategy based on $\ket{\psi_{AB}}$, with Alice flipping her result in case she has a skew symmetric observable, allows Alice and Bob to win with probability $1$.\\
\\
The general advantage of playing the doily game over the Mermin game, in each scenario, is the lower classical bound. This provides a greater chance of demonstrating quantum phenomena in a noisy quantum computer. Such a demonstration is made only when, implementing a given game $G$ and scenario $i$ as per Section \ref{sec:ibmtest} or otherwise, the computed success rate for the players, $\sigma_{i}(G)$, satisfies $\sigma_{i}(G) > \omega_{i}(G)$. \\

\subsection{More games based on larger configurations}
It is a natural question to then consider other point-line configurations that could be used for implementing a Mermin-like pseudo-telepathy game and compute their $\omega_i(G)$. For instance the Doily that encodes the commutation relations of the two qubit Pauli group can be generalized to a geometric structure denoted by $\mathcal{W}(2n-1,2)$ and called the symplectic polar space of rank $n$ and order $2$ that encode the commutation relation of the $n$-qubit Pauli group $\mathcal{P}_n$. In $\mathcal{W}(2n-1,2)$ the points correspond, up to a global phase, to non-trivial $n$-qubit Pauli observables and co-linear points correspond to mutually commuting observables. For example $\mathcal{W}(5,2)$ which encodes the commutation relations of $\mathcal{P}_3$ contains $63$ points and $315$ lines. These geometries have been studied in details in the context of quantum information in \cite{henri_contextuality_2022, muller_multi_qubit_2022}. It is in particular known that they are contextual \cite{cabello_proposed_2010,holweck_testing_2021}, and richer structures can provide in both scenarios a higher gap between the classical and quantum strategies.

\section{Implementing Quantum Games in a Noisy Quantum Computer}\label{sec:ibmtest}
Two basic methods of implementation for the given games and scenarios will be used - the unitary transform and the qubit delegation methods. Each will be outlined in this section, with the advantages and disadvantages of each provided. \\
We will use the circuit formalism to present our different implementations. All our codes and results can be found on \url{https://quantcert.github.io/quantum_game} and are implemented using Qiskit \cite{qiskit} the open-source software that allows to run programs on the IBM Quantum Experience \cite{ibmq}. 
In both scenarios Alice and Bob share the entangled states $\ket{\psi_{AB}}$ of Eq (\ref{eq:state}). In the circuit formalism that quantum state is implemented by:
\begin{center}
\begin{figure}[!h]
$\Qcircuit @C=1em @R=.7em {
\lstick{\text{Alice $1$st qubit}} & \gate{H} &\ctrl{2} & \qw &  \qw \\
\lstick{\text{Alice $2$nd qubit}}& \gate{H} & \qw & \ctrl{2} & \qw \\
\lstick{\text{Bob $1$st qubit}}& \qw & \targ & \qw & \qw \\
\lstick{\text{Bob $2$nd qubit}}& \qw & \qw & \targ & \qw
}$
\caption{Circuit implementing the entangled quantum state $\ket{\psi_{AB}}$.}
\label{fig:circuit_basics}
\end{figure}
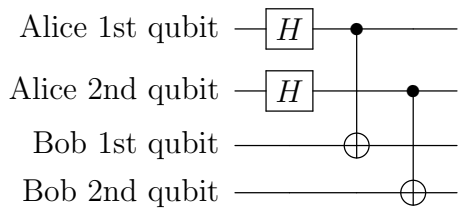
\end{center}
\begin{figure}[!h]
\begin{center}
$\Qcircuit @C=1em @R=.7em {
\lstick{\text{Alice $1$st qubit}} & \gate{H} &\ctrl{2} & \qw &  \qw & \multigate{1}{U_A} & \meter\\
\lstick{\text{Alice $2$nd qubit}}& \gate{H} & \qw & \ctrl{2} & \qw & \ghost{U_A} & \meter\\
\lstick{\text{Bob $1$st qubit}}& \qw & \targ & \qw & \qw & \multigate{1}{U_B} & \meter\\
\lstick{\text{Bob $2$nd qubit}}& \qw & \qw & \targ & \qw & \ghost{U_B} & \meter
}$
\caption{Circuit implementing the entangled quantum state $\ket{\psi_{AB}}$ followed by the unitary transformations that correspond to the Unitary Transform Method (Subsection \ref{sub:unitary}).}
\end{center}\label{fig:circuit_unitary}
\end{figure}
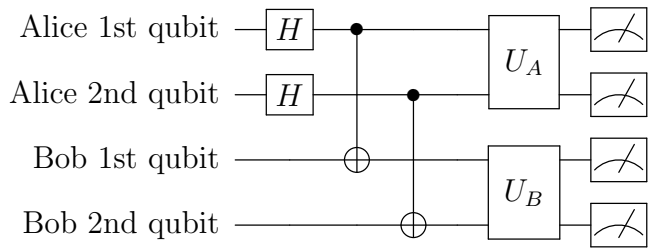
We will first outline the different implementation methods in general, and then specifically for each of the game scenarios in question.\\

\subsection{The Unitary Transform Method}\label{sub:unitary}
The unitary transform method utilises straightforward basis transformations for each of Alice and Bob's two-qubit system to transform them from the computational basis into the basis of mutual eigenvectors of their  respective contexts. We follow the description initially proposed in \cite{Brassard05}, who used it for the Mermin game - although it is equally applicable to the Doily game. \\
The method to compute a basis transformation matrix is as follows, for a given player and context:
\begin{enumerate}
    \item The spans of the eigenvectors are computed for each two-qubit operator on the context, associated with the eigenvalues $\pm 1$.
    \item The intersection of these spans are computed, for each combination of eigenvalues per operator. As the intersection is common to all three operators, this computation is only done for two of them.
    \item Each intersection is spanned by a single $4$ dimensional vector, giving 4 vectors for each combination of eigenvalues from the two operators.
    \item These four vectors form the common eigenvector basis matrix, whose conjugate transpose gives us the required transformation from the ``contextual" basis to the computational one.
\end{enumerate}
These unitary matrices are then applied to the circuit, one for Alice's two registers and one for Bob's. Measurements are then made on the respective qubits and registered onto 2 classical bits per player, representing the measurement results for the first two operators on the given context. The third bit is given by the parity conditions for that context - a negative parity representing an odd number of `$1$'s in the $3$-bit string representing the context measurement results.\\
For the Mermin game, we generated the above transformation matrices for each context and cycled through each of the 10 possible two-qubit Mermin grids for each game. As only two operators were ``used" in each computation, and different choices of operators resulted in different circuit implementations (resulting in different levels of noise), every possible ``reordering" of each grid was done for each game, representing different valid choices of which operators Alice and Bob would choose for their context measurements. \\
For the line-line scenario, each possible combination of intersecting rows and columns were given to Alice and Bob respectively, and measurements in the contextual bases were performed, generating the $3$-bit string for each player for each context. Success was then checked - if the bit in Alice's string at the same position as Bob's column number matched the corresponding bit in Bob's string in the Alice row position, then the players were given a success value of $1$, otherwise $0$. For each context this was run a total of $8162$ times in IBM's ``Lagos" backend. The overall average success for the best performing of the 10 grids, across each of the 9 questions per grid, was computed and given in Table \ref{tab:results_unitary}.\\
The above was completed similarly for the doily game, although there is only one unique doily arrangement containing all two-qubit operators (up to ordering, i.e. rotation). For each of the 15 intersection points, all possible choices of lines were given to Alice and Bob, and unitary matrices computed as before, and if the measurement results for each player at the ``intersection" point were consistent then a winning score of $1$ was assigned. This was similarly averaged out over the same number of computational attempts and all reorderings of the arrangement. The results are also given in Table \ref{tab:results_unitary}.\\
\\
The unitary transform method is singularly suitable for the \textit{line-line} scenario, as a context is given to each of Alice and Bob. However it has certain advantages and disadvantages which should be considered:
\\
\textbf{Advantages} 
\begin{itemize}
    \item General approach applicable to any context, regardless of arrangement.
    \item Requires no more qubit registers than the ones representing Alice and Bob's shared entangled states. 
\end{itemize}
\textbf{Disadvantages}
\begin{itemize}
    \item Computed unitary matrix depends on the operators involved in the context, so choice of operators results in different transforms.
    \item IBM machines manifest the unitary transformations often using many gates, resulting in considerable noise in the circuit dependent on the operators involved.
    \item Does not allow one to play the point-line (PL) scenario game (see Section \ref{sec:point_line_unitary}).
\end{itemize}
The second point in the disadvantages section is particularly salient here - the circuit, in order to implement the transformation matrices, introduces many auxiliary gates. Each gate introduces some amount of noise in the system itself, but this can be compounded by gates between unconnected qubits due to transpilation (see Section \ref{sec:simulators}). This noise generated by the matrices here reduces our computed success rate from $100\%$ down to a level below the classical bound - meaning no such quantum phenomena can be robustly said to exist in the process. There are two general ``fixes" for such an issue - find an alternative method that produces less noise (see Section {\ref{sec:delegation}), or find a game with a larger gap between classical and quantum winning bounds so that even with noise the success rate is higher than the classical bound. The doily line-line game provides the largest gap, however this is still not enough to overcome the issue with noise. \\

\subsection{The Delegation Method} \label{sec:delegation}
The delegation method is an alternative method for implementing a given quantum game, resulting more control over the final circuitry at the expense of using more qubit registers. It is also applicable to the \textit{point-line} scenario unlike the unitary method. \\
The circuit setup is as follows:
\begin{enumerate}
    \item For a given 1-qubit operator, the basis of the corresponding qubit is changed to match the  measurement basis (see Table \ref{tab:basis_change} for the different changes of basis).
    \item  The result of the measurement is encoded onto an auxiliary, called delegation register, via a CNOT gate. The qubit one wants to measure is then sent back to its original basis by the inverse change of basis.
    \item A measurement is then performed on this delegation qubit in the computational basis, giving a measurement result without destroying the original state. 
    \item The state can then be measured thanks to  other delegation qubits for secondary operators in a given context.
\end{enumerate}. 
\\
\begin{table}[h!]
\centering
\begin{tabular}{c|l}
    1-qubit operator & Gate(s) \\ \hline
    $Z$ & - \\
    $X$ & $H$ \\
    $Y$ & $S^{\dagger} H$
\end{tabular}
\caption{Basis change gates for a qubit register. Each register is initially in the computational $Z$-basis.}
\label{tab:basis_change}
\end{table}
\\
As we are concerned only with two-qubit measurement results and not the individual results of the 1-qubit operators, we can delegate measurements for both qubits of a player onto the same delegation qubit. These states are combined via bit-addition, giving a two-qubit state on a single qubit register, which can be measured directly. \\
\\
This method is illustrated with the following example of a point-line game. Alice and Bob are assigned two pairs of entangled qubits as before. Alice, having only one vertex and thus one measurement to perform, changes the bases of her qubits depending on the two-qubit operator $\mathcal{O}_{A}$ using Table \ref{tab:basis_change}.\\
This converts her qubits into the computational ($Z$) basis, on which she performs a direct measurement on each qubit. \\
As for Bob, he must now perform two measurements on his two-qubit state, corresponding to the first two operators $\mathcal{O}_{B,1}$, $\mathcal{O}_{B,2}$ in his context. The results of the third operator is deduced from the result of the above two and the parity ($\pm I_{4}$) of the context. This is not possible to do directly on his qubits as any measurement on one register will destroy the state for the following measurement. \\
Bob then delegates the state of his qubits onto a fifth qubit register which will be used to measure $\mathcal{O}_{B,1}$. He does this by first transforming the bases of his two qubits using the same gates as above, then performing a CNOT gate from his two qubits targeted onto this fifth register, then performing the inverse gates to return the initial qubits to their original $Z$ basis. The act of delegating both qubit measurements onto the one register performs an addition mod-$2$ operation on the qubit, meaning he can measure this delegation qubit directly now to get the overall value of $\mathcal{O}_{B,1}$. He then measures $\mathcal{O}_{B,2}$ directly on his original qubits as Alice did on hers. An example circuit demonstrating this method is given in Fig. \ref{fig:delegation_qubit}.

\begin{figure}[!h]
\begin{center}
$\Qcircuit @C=1em @R=.7em {
\lstick{\text{Alice $1$st qubit}} & \gate{H} & \ctrl{2} & \qw &  \qw & \gate{H} & \qw & \qw & \qw & \qw  & \meter  \\
\lstick{\text{Alice $2$nd qubit}}& \gate{H} & \qw & \ctrl{2} & \qw & \gate{S^\dag} & \gate{H} & \qw & \qw & \qw & \meter  \\
\lstick{\text{Bob $1$st qubit}}& \qw & \targ & \qw & \qw & \gate{S^\dag} & \gate{H} & \ctrl{2} & \gate{H} & \gate{S} & \meter \\
\lstick{\text{Bob $2$nd qubit}}& \qw & \qw & \targ & \qw & \gate{H} & \ctrl{1} & \qw & \gate{H} & \qw & \meter \\
\lstick{\text{Delegation qubit}}& \qw & \qw & \qw & \qw & \qw & \targ & \targ & \qw & \qw & \meter \\
}$
\caption{Circuit implementing the delegation method for the operators $\mathcal{O}_{A} = XY$, $\mathcal{O}_{B,1} = YX$, $\mathcal{O}_{B,2} = ZZ$.}
\end{center}
\end{figure}
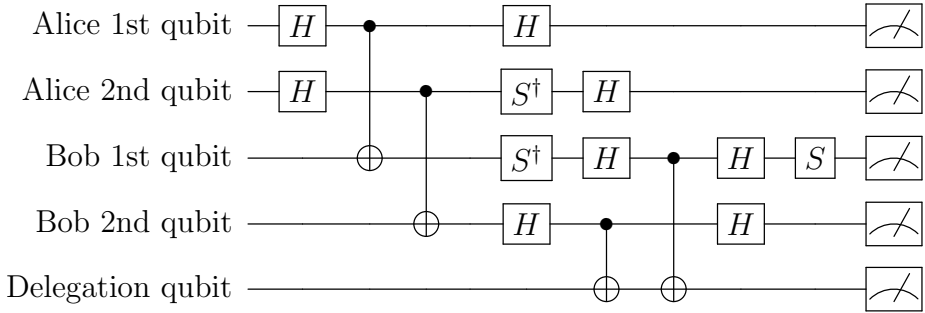\label{fig:delegation_qubit}

To reduce noise as much as possible, the particular topology of the quantum machine must be taken into account. In this example, using the ``Lagos" backend, the topology is given in Figure \ref{fig:jakarta_top}. Thus, we associate registers $0, 4$ with Alice's qubits, $1, 5$ with Bob's initial qubits, and register $3$ with the delegated qubit. This ensures only direct CNOT gates are used. \\

\begin{figure}[!h]
 \includegraphics[width=5cm]{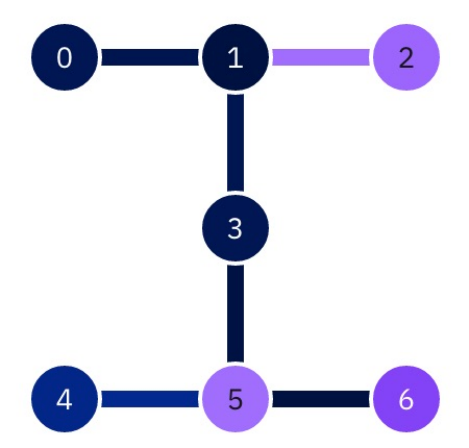}
 \caption{The CNOT topology for IBM ``Lagos" backend. Nodes represent qubits, and edges represent direct CNOT connections. Credit: IBM, https://quantum-computing.ibm.com, 2023}\label{fig:jakarta_top}
\end{figure}

For the point-line scenario, as only one player is given a context and thus must make additional measurements, only one extra delegation qubit is required. For the line-line scenario, one needs 2 auxiliary qubits to retain the value of the first measurement in each player's context. Then the second measurements can take place on the original qubits, with the third determined from the parity of the lines.
\\
Like the unitary transform method, the delegation method comes with certain advantages and disadvantages:
\\
\textbf{Advantages} 
\begin{itemize}
    \item Does not require the computation of unitary matrices, the method depends only on reading the operators involved and implementing the associated gates.
    \item Greater control of which gates are needed in basis changes compared to unitary method, which can reduce noise via fewer gates.
    \item With better knowledge as to which qubit registers require CNOT gates between them, noise can be further reduced by associating registers to players based on a given machine's CNOT topology.
    \item Allows one to play the point-line (PL) scenario (see Section \ref{sec:point_line_unitary}).
\end{itemize}
\textbf{Disadvantages}
\begin{itemize}
    \item Requires larger qubit circuits than the unitary method. 5 qubits total required for the point-line scenario, 6 for the line-line scenario.
\end{itemize}
The requirement of additional qubits is not a major roadblock for examining the line-line scenario, as currently IBM allows free access to 7-qubit machines for research purposes. This becomes a bigger issue with games involving more qubits, for example a two player game involving 3-qubit operators in each context would require a total of 8 qubits when using the delegation method.\\
\\
This method was used for the Mermin and doily point-line and line-line games, with results given in Table \ref{tab:results_delegated}.

\subsection{Unitary Method \& Point-Line Scenario}\label{sec:point_line_unitary}
As mentioned, the unitary method does not allow for the point-line scenario to be played. In this section we will explain why that is the case. 
\\
In every game described so far, the players share the state $|\psi_{AB}\rangle$ described in  \eqref{eq:state}. For the delegation method, players perform basis changes and measurements based on the operators $\mathcal{O}^{A}_{i}, \mathcal{O}^{B}_{i}$ in their contexts (or individual $\mathcal{O}^{A}$ for the point-line game). When the operators are equal, i.e. at the point of intersection of the geometry, then the state $|\psi_{AB}\rangle$ is an eigenstate of the joint measurement and satisfies \eqref{eq:same_eigvalue_sym} with eigenvalue $-1$ for skew operators and $+1$ otherwise. This, along with the skew-flipping protocol, ensures that the measurement outcome for the same operator is obtained for both the line-line and point-line  scenarios.
\\
For the unitary method, one performs the unitary transformations and obtains the state ${U}_{A} \otimes {U}_{B} |\psi_{AB}\rangle$. While $|\psi_{AB}\rangle$ is not an eigenstate of this operator, it does ensure that the bit value obtained in the extended bit outputs match on the intersection position, ensuring victory in the game. This works for the line-line scenario where both players must change basis based on two of the 2-qubit operators in their contexts.
\\
Attempting to play the point-line scenario using the unitary method produces a state $\mathcal{O}^{A} \otimes {U}_{B} |\psi_{AB}\rangle$. The state $|\psi_{AB}\rangle$ is neither an eigenstate of this operator nor does it ensure agreement on the intersection-positioned bits in the extended output. Thus, the unitary method is not appropriate for the point-line scenario, and it has not been applied as such in this work.

\subsection{Simulators \& Noise}\label{sec:simulators}
The IBM Quantum Experience allows for running quantum circuits on simulated quantum computers, including simulated noise models based on a given backend. In this work we utilise this function to demonstrate the validity of our circuits and the quantum advantages of the associated games. 
\\~\\
However one should be wary of the results given by simulated noisy models when the circuit is sufficiently complex. The noise models are based on the CNOT and readout error rates of the individual qubits and CNOT links in the given backend. But when a circuit is implemented into a NISQ computer, the circuit design undergoes transpilation, adding necessary gates to connect qubits via CNOTs and other 2-qubit gates that don't have a direct connection in the backend topology (see Fig. \ref{fig:jakarta_top}). Such transpilation (and additional gates) are not taken into account by a simulator with given noise model.
\\
In the line-line games under investigation here, all circuits include at least one qubit-qubit connection that is not given by a direct link in the topology. That means that the transpilation process adds multiple additional 2-qubit gates that the simulator does not register. Thus the noise levels predicted by the simulators are generally lower than the actual noise levels seen in the machines, and the success rate does not beat the classical bound for any of our experiments, despite suggestive figures from the simulated runs.
\\
In the point-line scenario, the simulated noise more closely approximates the observed levels, but the higher classical bound proves too large for all apart from the noiseless simulators.
\\
\section{Results}\label{sec:results}

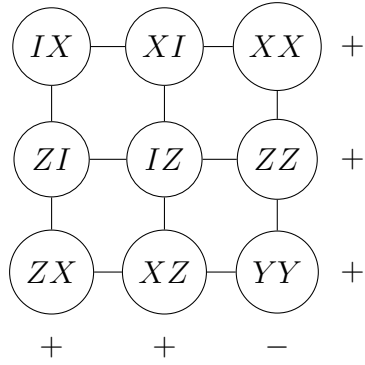
\begin{figure}[h!]
\centering
\begin{tikzpicture}[
operator/.style={circle, draw, minimum size=1.0cm},
]
\foreach \x/\y/\num in {1/3/$IX$, 2/3/$XI$, 3/3/$XX$, 1/2/$ZI$, 2/2/$IZ$, 3/2/$ZZ$, 1/1/$ZX$, 2/1/$XZ$, 3/1/$YY$}
    \node[operator] (\x-\y) at (1.5*\x, 1.5*\y) {\num};
    
\draw[] (1-1.east) -- (2-1.west);
\draw[] (2-1.east) -- (3-1.west);
\draw[] (1-2.east) -- (2-2.west);
\draw[] (2-2.east) -- (3-2.west);
\draw[] (1-3.east) -- (2-3.west);
\draw[] (2-3.east) -- (3-3.west);

\draw[] (1-1.north) -- (1-2.south);
\draw[] (1-2.north) -- (1-3.south);
\draw[] (2-1.north) -- (2-2.south);
\draw[] (2-2.north) -- (2-3.south);
\draw[] (3-1.north) -- (3-2.south);
\draw[] (3-2.north) -- (3-3.south);

  \foreach \y in {1,2,3}
    \node at (5.5, 1.5*\y) {$+$};
    
    \node at (1.5*1, 0.5) {$+$};
    \node at (1.5*2, 0.5) {$+$};
    \node at (1.5*3, 0.5) {$-$};

\end{tikzpicture}
\caption{The particular grid used for testing the Mermin game in the LL and PL (D) scenario. Operators along a context multiply to give the identity, scaled by $\pm1$, the sign of which is indicated at each row and column. In this version, Bob returns an array $\vec{b}=(b_{1},b_{2},b_{3})$ whose elements multiply to $+1$ for $y=1,2$ and multiply to $-1$ for $y=3$.}\label{fig:mermintest}
\end{figure}

Tables \ref{tab:results_unitary} and \ref{tab:results_delegated} for the unitary and delegation methods, respectively, show the results on the IBM Lagos machine. Results are given for both the line-line scenarios and the point-line scenario for the delegation method. For each instance of the Doily game, every game was ran with $8192$ shots, over 6 permutations of operator ordering for each line, for a total of $8192 \times 6 \times 15 \times 3 = 2,211,840$ shots per result. For the Mermin figures, each of the 10 Mermin grids were extracted from the Doily geometry, and the best performing one (Fig. \ref{fig:mermintest}) was used.

\newcolumntype{P}[1]{>{\centering\arraybackslash}p{#1}}
\renewcommand{\arraystretch}{1.3}

\begin{table}[h!]
\begin{tabularx}{15.57cm}{|P{1.4cm}P{1.5cm}P{2cm}||P{2cm}P{2cm}P{2cm}||P{2cm}|} \hline 
     \qquad \newline \text{Game} \newline $G$ & \qquad \newline \text{Scenario} \newline $i$ & \qquad \newline Date & Noiseless Simulation (U) & Noisy Simulation (U) &  \qquad \newline $\sigma_{i}(G)$ & \qquad \newline $\omega_{i}(G)$ \\ \hline \hline
    \multirow{2}{*}{\text{Mermin}} & \text{LL} & \text{Feb 2023} & $100\%$ & $93.637\%$ & \boldsymbol{$85.624\%$} & $88.\overline{8}\%$ \\ 
     & \text{PL} & $-$ & $-$ & $-$ & $-$ & $94.\overline{4}\%$ \\ \hline
    \multirow{2}{*}{\text{Doily}} & \text{LL} & \text{Feb 2023} & $100\%$ & $96.444\%$ & \boldsymbol{$82.889\%$} & $86.\overline{6}\%$ \\ 
     & \text{PL} & $-$ & $-$ & $-$ & $-$ & $93.\overline{3}\%$ \\ \hline
\end{tabularx}
\caption{Results for computing the observed success rate $\sigma_{i}(G)$ on the Lagos NISQ Computer, using the unitary method (U) for both the Mermin and Doily games on the given dates. Shown are results for the line-line (LL) scenario, as the point-line is not implementable by (U). Results are listed alongside their noiseless and noisy simulated successes. No values of $\sigma_i(G)$ beat the classical bounds $\omega_{i}(G)$ shown, due to noise in the transpilated circuits. The simulated values show promise for the line-line games, even with noise models.} 
\label{tab:results_unitary}
\end{table}

\begin{table}[h!]
\begin{tabularx}{15.57cm}{|P{1.4cm}P{1.5cm}P{2cm}||P{2cm}P{2cm}P{2cm}||P{2cm}|} 
    \hline
    \qquad \newline \text{Game} \newline $G$ & \qquad \newline \text{Scenario} \newline $i$ & \qquad \newline Date & Noiseless Simulation (D) & Noisy Simulation (D) & \qquad \newline $\sigma_i(G)$ & \qquad \newline $\omega_i(G)$ \\ \hline \hline
    \multirow{2}{*}{\text{Mermin}} & \text{LL} & \text{Aug 2023} & $100\%$ & $92.948\%$ & \boldsymbol{$86.849\%$} & $88.\overline{8}\%$ \\ 
     & \text{PL} & \text{Sept 2023} & $100\%$ & $93.646\%$ & \boldsymbol{$91.781\%$} & $94.\overline{4}\%$ \\ \hline
    \multirow{2}{*}{\text{Doily}} & \text{LL} & \text{Aug 2023} & $100\%$ & $92.106\%$ & \boldsymbol{$82.021\%$} & $86.\overline{6}\%$ \\ 
     & \text{PL} & \text{Sept 2023} & $100\%$ & $92.935\%$ & \boldsymbol{$90.447\%$} & $93.\overline{3}\%$ \\ \hline
\end{tabularx}

\caption{Similar result table to Table \ref{tab:results_unitary}, but for the delegation method (D). Included are both line-line (LL) and point-line (PL) scenarios. Again, simulated results show promise, but circuit transpilation generates too much noise in the Lagos results $\sigma_i(G)$. The bounds are identical for both methods.} 
\label{tab:results_delegated}
\end{table}

When run on simulators without noise, all games, methods and scenarios resulted in winning with probability 1, clearly demonstrating the theoretical quantum advantage over the classical picture and the validity of both implementations (scenarios). When run with noise models, the results still beat out the classical bound for the line-line scenario, for both games and methods. This demonstrates that for a quantum backend with an appropriate topology, both Mermin and Doily games can demonstrate  quantum advantage even with similar per-gate noise levels as in the IBM Lagos machine.\\

The observed success rates $\sigma_i$ of the Mermin and Doily games as shown in Tables \ref{tab:results_unitary} and \ref{tab:results_delegated} do not beat the classical bounds. 

As noted the simulators do not transpile the circuits to match the topology of the given machine, and so auxiliary gate effects are excluded from the simulated results. That should mostly explain the difference between the noisy simulation and the actual computation on the quantum computer. However, despite the fact that no classical bound is beaten, the results are close to exhibit non classical 
behavior. 
There is no real difference in observed success rate for the line-line game between the two methods used. In contrast, there is a sizable gap in performance using the noisy simulator, with the unitary method outperforming the delegation one. Indeed the delegation method implies more qubits by definition and that could explain the notable difference observed with the noisy simulator. However the observed success rates indicate that it may not be the case on real quantum computer. Maybe here also the noisy simulation does not fully take into account the extra added gates from the transpilation process and in this respect the delegation method can exhibit an advantage.

\section{Conclusion}\label{sec:conclusion}
In this paper we discuss the implementation of quantum games based on contextual configuration of observables. We started with the well-known Mermin game and its implementation as described in \cite{brassard_quantum_2005}. We also provided a new way of implementing these type of game, the delegation method, as well as a new type of game based on another contextual configurations made of $15$ observables and $15$ contexts. This richer game, built on this configuration, exhibits a bigger gap between the probability of winning it with a classical or quantum strategy. Finally one implements theses games on an online quantum computer.

One believes that the study of quantum games is of importance for the development of quantum information \cite{bravyi2020quantum}, especially in the context of NISQ computers. These games provide a current use-case of existing NISQ computers as a quantum laboratory, even with the moderate noise levels therein. We also think that the range of applications of such types of games, once the quantum advantage is repeatedly established on NISQ computers, deserve further investigation. 

Our geometrical approach also invites us to consider more advanced geometrical structures. Here one only considered observable made of two qubit operators but the use of three-qubit observables would allow us to explore richer finite geometric configurations. This will be done in a future work.
\section*{Acknowledgment}
This work is supported by the Graduate school EIPHI (contract ANR-17-EURE-
0002) through the project TACTICQ. We acknowledge the use of the IBM Quantum 
Experience and the IBMQ-research program. The views expressed are those of the authors and do not reflect the official policy or position of IBM or the IBM Quantum Experience team. One would like to thank the developers of the open-source framework Qiskit.

\pagebreak

\printbibliography
\end{document}